\documentclass[conference,english]{IEEEtran}
\usepackage[utf8]{inputenc}
\usepackage{balance}
\usepackage{babel} 
\usepackage{graphicx}
\usepackage{subfigure}
\usepackage{color}
\usepackage{amsmath}
\usepackage{multirow}
\usepackage{url}
\usepackage{tikz}
\usepackage[colorlinks,bookmarksopen,bookmarksnumbered,citecolor=red,urlcolor=red]{hyperref}

\newcommand\copyrighttext{%
  \footnotesize \textcopyright 2009 IEEE. Personal use of this material is permitted.
  Permission from IEEE must be obtained for all other uses, in any current or future 
  media, including reprinting/republishing this material for advertising or promotional 
  purposes, creating new collective works, for resale or redistribution to servers or 
  lists, or reuse of any copyrighted component of this work in other works. 
  DOI: \href{http://dx.doi.org/10.1109/GLOCOM.2009.5425710}{10.1109/GLOCOM.2009.5425710}
}
\newcommand\copyrightnotice{%
\begin{tikzpicture}[remember picture,overlay]
\node[anchor=south,yshift=10pt] at (current page.south) {\fbox{\parbox{\dimexpr\textwidth-\fboxsep-\fboxrule\relax}{\copyrighttext}}};
\end{tikzpicture}%
}

\title{Improved Opportunistic Sleeping Algorithms for LAN Switches}

\author{%
  \authorblockN{M. Rodríguez-Pérez,
    S.~Herrería-Alonso,
    M.~Fernández-Veiga,
    C.~López-García}
  \authorblockA{Dept.\ of Telematics Engineering\\
    ETSE Telecomunicación\\
    Campus universitario s/n \\
    36310 Vigo, Spain}
}

\begin{document}
\maketitle
\copyrightnotice
\begin{abstract}
  Network interfaces in most LAN computing devices are usually severely
  under-utilized, wasting energy while waiting for new packets to arrive. In
  this paper, we present two algorithms for opportunistically powering down
  unused network interfaces in order to save some of that wasted energy. We
  compare our proposals to the best known opportunistic method, and show that
  they provide much greater power savings inflicting even lower delays to
  Internet traffic.
\end{abstract}

\section{Introduction}
\label{sec:introduction}

The total amount of energy needed to power networking infrastructure has been
rising as more devices have been getting connected to the network. Moreover,
the nominal link capacities have also been growing, demanding more and more
power for the card transceiver with each speed increase. In the last few
years, these greater power demands have coincided with increasing
environmental concerns and higher operating costs associated with networking
equipment.

Traditionally, the design of networking equipment concentrated in maximizing
performance, irrespectively of power demands. However, as the operating costs
associated with heat dissipation and energy consumption continue to increase,
this trend is starting to reverse. This is not a big surprise, as other
computer related fields, like computer processors, graphic cards, ..., also
suffered this change in optimization focus not too long ago.

Many of the places where energy could be saved in the current Internet design
were first identified in~\cite{gupta03:_green_of_inter}. One of those places
are the links where actual transmission takes place. Until recently, only
power needs of mobile devices were devoted some
consideration~\cite{ye02:_energ_effic_mac_protoc_for}. However, energy
consumption in wired mediums cannot be neglected, when modern gigabyte cards
already demand around $4\,$W and, soon to be the norm, 10 gigabyte cards
consume in the order of tens of
watts~\cite{gupta07:_using_low_power_modes_for,zhang08:_real_time_perfor_analy_of,patel-predd08:_energ_effic_ether}.
At the same time, most network interfaces sit unused most of the time wasting
too much power~\cite{gupta04:_feasib_study_for_power_manag}. For instance,
even in highly-utilized backbone switches averaged traffic loads below
$30\,\%$ are normal~\cite{jardosh07:_towar_energ_star_wlan_infras}. This is
not only a problem to energy constrained devices, like laptops, but also a
source of noticeable amounts of unnecessary heat for under-utilized switches.

In this paper we build on an opportunistic sleeping algorithm for powering
down network interfaces when there is low probability of buffer overruns while
the interface is down~\cite{gupta07:_using_low_power_modes_for}. We analyze
its shortcomings and propose several enhancements to the algorithm that
greatly augment its power savings. Then, we simplify the resulting algorithm
the most so as to increase the chances making it deployable in cheap hardware
too. The result is an algorithm much simpler than the original, and that is
able to achieve much higher energy savings with less added delay and a
negligible increase of packet losses.

The rest of this paper is organized as follows. Section~\ref{sec:power-model}
shows the model on which we will build on. Our proposals are described and
compared against the original
Gupta-Singh~\cite{gupta07:_using_low_power_modes_for} algorithm in
Section~\ref{sec:opport-sleep}. In Section~\ref{sec:evaluation} we present an
evaluation of the performance of the different algorithms. Finally, our
conclusions are laid out in Section~\ref{sec:conclusions}.

\section{Line Card Power Model}
\label{sec:power-model}

In this paper we assume that line cards and, more precisely, its individual
interfaces can be put to sleep. This is in accordance with previous works in
the
subject~\cite{gupta07:_using_low_power_modes_for,gupta04:_feasib_study_for_power_manag,nedevschi08:_reduc_networ_energ_consum_via}
and the general trend in other related fields, such as in embedded devices or
in desktop computers, where different system parts can be powered down at will
to save power.

Network interfaces could theoretically offer a fine grained control over what
parts of the hardware are active, their operating speeds, etc. to adjust power
consumption to our needs. This control could be exposed to the operating
system via different \emph{sleeping profiles} akin to those presented by
ACPI~\cite{06:_advan_config_and_power_inter_specif}. We restrict ourselves to
a simpler design with just one sleeping state where the network card is
completely shut down without even the ability to sense the line for incoming
traffic. This permits maximum savings for the interface receiving part.

We will thus consider four operating states with different associated power
profiles for a network card: one sleeping state, two awake states and a
transitioning state from sleeping to active. The two awake states differ in
the use of the transceiver port. This port can be either \emph{active,} when
the card is transmitting or \emph{idle,} if the card is not doing useful work.
Although we could define more than one transitioning state we ignore all but
the one from sleeping to awake. All the other transitions can take place
almost instantaneously and without incurring in additional power consumption,
but the transition from sleeping to awake needs that the clocks of the two
link endpoints network cards be resynchronized, and this takes some time and
needs as much power as that needed for active transmission, so we cannot
simply ignore it.\footnote{The resynchronization phase involves, in effect,
  the transmission of known signals to properly adjust the clocks. This
  transmission length is by no means negligible, and also consumes a non
  trivial amount of power.}

We therefore define the following power consumption vector for a given
interface: $\vec p = \{p_\mathrm{a}, p_\mathrm{i}, p_\mathrm{s}\}$, where
$p_\mathrm{a}$ is the power consumption when the interface is actively
transmitting data, $p_\mathrm{i}$ is the power drain when the interface is
awake, but otherwise idle and, finally, $p_\mathrm{s}$ is the (small) power
drain while the interface is sleeping. Recall that $p_{\mathrm{a}}$ is also
the power needed while transitioning from sleeping to awake.

\section{Opportunistic Sleeping}
\label{sec:opport-sleep}

For an opportunistic sleeping algorithm to be defined, three are the questions
that must be answered:
\begin{itemize}
\item When to sleep
\item How long to sleep
\item When to wake up
\end{itemize}

Throughout this section we will first describe the algorithm proposed by Gupta
and Singh in~\cite{gupta07:_using_low_power_modes_for} and then we will
present our enhancements to it with their motivations.

\subsection{The Gupta-Singh Algorithm}
\label{sec:gupta-singh-algor}

In~\cite{gupta07:_using_low_power_modes_for} the authors present two related
algorithms that give answer to the three aforementioned questions. We will
concentrate only on the second algorithm one, as it provides greater savings
than the first.

Their method relies on the assumption that packet arrivals to an Internet
queue follow a Poisson distribution in small
timeframes~\cite{karagiannis04:_nonst_poiss_view_of_inter_traff}. So they
employ the average inter-arrival time of the last few packets ($5$ in their
implementation) to obtain a rough estimation of the arrival process rate
($\lambda$) in the short term.

When the queue occupation ($q$) goes below a certain threshold $b$, this
information is used to estimate the time that the interface can be put to
sleep ($t_{\mathrm{s}}$) without risking that the buffer occupation goes above
$b$. For this, the random variable $X_k$ is defined as the sum of $k$
independent and identically distributed exponential distributions with rate
$\lambda$, where $k$ is the spare capacity in the queue below $b$, that is
$k=b-q$. In effect, $X_k$ is just an Erlang-$k(\lambda)$ distribution, and the
sleeping time is calculated so that
\begin{equation}
  \label{eq:erlang-k-incognito}
  P\left(X_k \ge t_{\mathrm{s}}\right) \ge 0.9.
\end{equation}
That is, there is a relatively small chance (10\%) that the queue occupation
will grow above $b$ while sleeping. In their paper they propose to calculate
$b$ as a small fraction of the total transmission queue size ($B$), for
example $b=0.1B$.

If $t_{\mathrm{s}}$ results to be greater than the transition time
($t_{\delta}$) the interface is put to sleep for $\max(t_{\mathrm{s}} -
t_{\delta}, t_{\max})$, with $t_{\max}$ being a configuration parameter. At
the same time, the sender communicates this value to the receiving interface
so that it can also enter the sleeping mode.

Once the sleeping timer fires, the interface resumes normal activity unless
the queue happens to be completely empty ($q = 0$). In this case,
a new sleeping interval of the same length is started. The receiving interface
notices this new sleeping interval as it senses the line when its own sleeping
timer expires and, as it measures no power it infers that a new sleeping
interval has started and returns to the sleeping state.\footnote{If the
  receiving card can sense the line while sleeping the last two questions, how
  long to sleep and when to wake up, converge into one. Whenever the upstream
  interface decides to recommence the transmission the receiver notices that
  there is again power in the line and wakes up. On the other hand, if the
  receiving card does not have this capability, the sender must restrict
  itself to only transmit when there is a change that the receiver can detect
  it. That is, when its sleeping timer expires.}

\begin{figure}
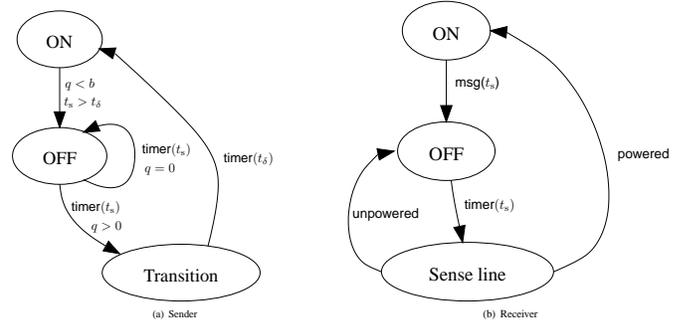

  \centering
  \resizebox{0.49\columnwidth}{!}{\subfigure[Sender]{\input{sender-orig.pstex_t}}}
  \resizebox{0.49\columnwidth}{!}{\subfigure[Receiver]{\input{receiver.pstex_t}}}
  \caption{The Gupta-Singh algorithm}
  \label{fig:gupta-orig}
\end{figure}

A concise drawing depicting the state diagram of the algorithm is represented
in Fig.~\ref{fig:gupta-orig}.

\subsection{Enhanced Gupta-Singh}
\label{sec:our-proposal}

We found some shortcomings in the previous algorithm that motivated us to try
to improve it and measure the differences.

For instance, the Gupta-Singh algorithm introduces unneeded delays when it
decides to sleep if there is still traffic in the queue. There is no reason
for not postponing the sleeping interval until there is no more traffic in the
queue, as all the time needed for emptying the queue will be used for
profitable work. Furthermore, if during this time, the incoming rate remains
low enough so that the queue finally drains, the new sleeping time will
probably be larger, as there will be more room in the queue to allocate
packets while sleeping. Recall also that the system losses work every time it
transitions to active state as there is a period of activity, the transition
time, in which no useful work is done, but the power consumption is high. So
it is more profitable to sleep once a long interval than several short
ones.

With this simple change, only sleeping when the queue is completely empty, we
were able to vastly improve the total sleeping time of the interfaces.
Moreover, the computation complexity needed for calculating $t_{\mathrm{s}}$
is greatly reduced.

In general, for solving $t_{\mathrm{s}}$ in~\eqref{eq:erlang-k-incognito} the
following equation must be solved
\begin{equation}
  \label{eq:erlang-k-expanded}
  1-\sum_{n=0}^k \mathrm{e}^{-\lambda t_{\mathrm{s}}} 
  \frac{(\lambda t_{\mathrm{s}})^n}{n!} \ge 0.9,
\end{equation}
$k$ being the number of packets that we can accommodate while sleeping. Sadly,
there exists no closed form formula for $t_{\mathrm{s}}$ and we must resort to
numerical approximations. In fact the Newton method gives good results in just
a few iterations taking $k \lambda^{-1}$ as the starting point. However, it is
not practical for the network interface to solve this equation every time
$\lambda$ or $k$ changes, that is after each packet arrival or departure.

The good news is that $P(X_k \ge t_{\mathrm{s}}) = \mathrm{f}(\lambda
t_{\mathrm{s}})$, as a fast inspection of eq.~\eqref{eq:erlang-k-expanded}
soon reveals. So, for any fixed value of $k$ the relation between
$t_{\mathrm{s}}$ and $\lambda$ becomes linear. With our simpler approach, that
only sleeps when the buffer is empty, $k$ remains constant (in fact, $k=b$).
Therefore the network card operator can pre-load any pair $(\lambda,
t_{\mathrm{s}})$ adequate to the buffer size and the card itself can easily
extrapolate values for different packet incoming rates.

A second shortcoming is that the Gupta-Singh algorithm tries to maximize the
total sleeping time as a way to maximize power savings. While both concepts
are greatly correlated, sometimes putting an interface to sleep has
predictable bad consequences in power consumption. There is certainly a
minimum sleeping interval, below which it is not profitable to sleep. Quite
the contrary, recall that once the sleeping interval finishes, the interface
goes through a transitioning phase when power consumption is like that of
active transmission. So, the total energy consumed during the sleeping
interval becomes $(t_{\mathrm{s}} - t_{\delta})p_{\mathrm{s}} + t_{\delta}
p_{\mathrm{a}}$. Before deciding to sleep, this quantity must be compared
against the energy consumed if the interface was awake, but idle, for the
sleeping interval duration. Thus
\begin{equation*}
  (t_{\mathrm{s}}-t_{\delta}) p_{\mathrm{s}} + t_{\delta} p_{\mathrm{a}} < t_{\mathrm{s}}
  p_{\mathrm{i}}
\end{equation*}
or, in a more direct form,
\begin{equation}
  \label{eq:profitable-sleeping}
  t_{\mathrm{s}} > t_{\delta}\frac{p_{\mathrm{a}}-p_{\mathrm{s}}}{p_{\mathrm{i}} - p_{\mathrm{s}}},
\end{equation}
becomes a necessary condition for a worthy sleeping interval. In the
evaluation section we will show how this seemingly trivial change can have a
dramatic effect in total energy savings.

\begin{figure}
  \centering
  \input{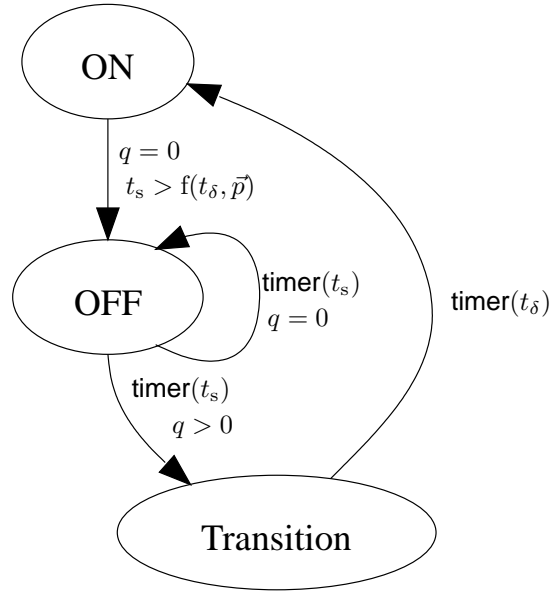}
  \caption{Our modified sender algorithm}
  \label{fig:gupta-improved}
\end{figure}

With the above mentioned changes the new state diagram for the sender looks
like the one represented in Fig.~\ref{fig:gupta-improved}. The receiver
algorithm remains unchanged.

\subsection{A More Streamlined Proposal}
\label{sec:parvo}

After performing the previous adjustments to the pristine Gupta-Singh
algorithm we tried to simplify even more the algorithm to improve its chances
of being deployed. 

For instance, is it possible to just try to sleep every time the queue gets
empty? After all, our modifications demanded this condition to be met and then
calculated an estimation of the maximum sleeping time based on the short time
incoming rate. If this rate was low enough, then the calculated $t_{\mathrm{s}}$
would be high enough to sleep. In hindsight this should be usually the case,
as the queue drains for a reason: the incoming rate is low. So, in our
simplified proposal, the interface sleeps every time the queue empties.

Second question is how long to sleep. For this,
eq.~\eqref{eq:profitable-sleeping} already provides a lower bound. It is not
worth to sleep less than the minimum provided by
eq.~\eqref{eq:profitable-sleeping}. We take this minimum as the sleeping
interval.

The final question is when to wake up the interface. Every time the sleep
timer fires the upstream interface measures its transmission queue length.
Although at first sight it may seem that the interface should transition to
active whenever the queue is not empty, this is not very sensible, because if
$q$ is too small it will be put to sleep again in a too short time, making the
transition to active and back to sleep unprofitable. It is better to queue
some traffic so that the costs associated with bringing the interface back to
active are small compared to the cost of transmitting the queued
packets.\footnote{Obviously, there should be some upper bound to the time
  spent sleeping with queued traffic so as to prevent starvation in the queue
  and too big delays.} With these considerations the minimum queue length for
waking up ($q_{\mathrm{w}}$) must meet that
\begin{equation}
  \label{eq:qw-min-implicit}
  p_{\mathrm{a}} \frac{q_{\mathrm{w}}}{C} = p_{\mathrm{a}} t_{\delta},
\end{equation}
that is
\begin{equation}
  \label{eq:qw-min-explicit}
  q_{\mathrm{w}} = C t_{\delta},
\end{equation}
where $C$ is the nominal interface bandwidth.

This condition also helps us to give suggestions about the minimum
transmission buffer size ($B$), as it must be big enough to hold at least
$q_{\mathrm{w}}$ packets while sleeping. A good approximation can be to make
$B$ an order of magnitude higher than $q_{\mathrm{w}}$. This way, the chance
of overflowing the buffer capacity while sleeping is diminished.

\begin{figure}
  \centering
  \input{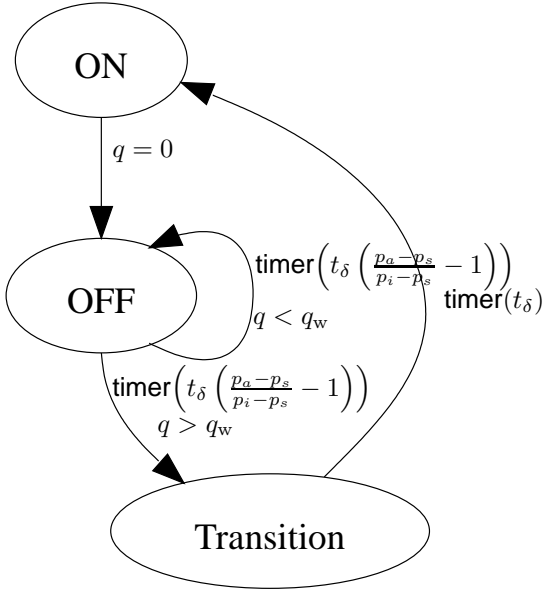}
  \caption{Streamlined proposal}
  \label{fig:dumb}
\end{figure}

The final state diagram for this simplified algorithm is represented in
Fig.~\ref{fig:dumb}.
Additionally, in table~\ref{tab:comparison} we show a summary of the
conditions used by the three described methods.
\begin{table*}
  \centering
  \begin{tabular}{l|c|c|c|}
    & \textbf{Gupta-Singh} & \textbf{Enhanced Gupta-Singh} &
    \textbf{Streamlined Algorithm}\\\hline

    \multirow{3}{*}{\emph{Sleep Condition}} &
    $q < b$ & $q = 0$ & \\
    &
    $t_{\mathrm{s}} | P\left(X_{b-q} \ge t_{\mathrm{s}}\right) \ge
    0.9$ &
    $t_{\mathrm{s}} | P\left(X_b \ge t_{\mathrm{s}}\right) \ge
    0.9$ & $q = 0$  \\ &
      $t_{\mathrm{s}} > t_{\delta}$ &
      $t_{\mathrm{s}} \ge
      t_{\delta}\frac{p_{\mathrm{a}}-p_{\mathrm{s}}}{p_{\mathrm{i}} -
        p_{\mathrm{s}}}$ & \\ \hline
      
      \emph{Sleep Interval} & $\max\{t_{\mathrm{s}} - t_{\delta},
      t_{\max}\}$ & $\max\{t_{\mathrm{s}} - t_{\delta},
      t_{\max}\}$ & $\max \left\{t_{\delta}\left(\frac{p_{\mathrm{a}}-p_{\mathrm{s}}}{p_{\mathrm{i}} -
          p_{\mathrm{s}}} - 1 \right), t_{\max} \right\}$ \\\hline
      
    \emph{Wake-up Condition} & $q > 0$ & $q > 0$
    & $q > q_{\mathrm{w}}$ \\\hline
  \end{tabular}
  \caption{Comparison between the three proposed sleeping algorithms.}
  \label{tab:comparison}
\end{table*}

\section{Evaluation}
\label{sec:evaluation}

We performed via simulation a comparison between our proposed algorithms and
the original Gupta-Singh proposal. For this we have employed the same dataset
that they used in their original
paper~\cite{gupta07:_using_low_power_modes_for}, which they kindly provided to
us. The data consist on the arrival times of packet in their internal network.
Sadly, the data lacks packet sizes, so, for our study we have decided to
initially consider a constant packet size of $1\,000\,$bytes. We then made use
of the ns-2 simulator to test our sleeping procedures in a gigabyte
link~\cite{ns-2}. For the sake of space we will only present here the results
for two of the data traces: the one with most activity (labeled \emph{High} in
the following figures) with an occupation factor $\rho=7.2\%$, and,
correspondingly, the one with the least (\emph{Low}), with $\rho=0.13\%$.

For the power vector we use the same values as
in~\cite{gupta04:_feasib_study_for_power_manag}, that estimated these values
extrapolating data from the wireless domain. That is, $p_\mathrm a=2\,$W,
$p_\mathrm i=1\,$W and $p_\mathrm s=0.1\,$W. In any case, the difference
between $p_\mathrm a$ and $p_\mathrm i$ matches that provided by some network
equipment providers~\cite{08:_nortely}.

For the rest of the parameters, we decided to take the same values as
in~\cite{gupta07:_using_low_power_modes_for} to provide the fairest
comparison. So that $t_\delta=0.5\,$ms and $t_{\max} = 2.5\,$ms. 
We also set $b=0.1B$ for both the original Gupta-Singh algorithm and our
modified proposal.
Finally, the
minimum profitable sleeping interval for our modified algorithms, is then
$t_\delta\frac{p_\mathrm a - p_\mathrm s}{p_\mathrm i-p_\mathrm s} =
1.06\,$ms. The only remaining parameter, needed for the streamlined algorithm
is $q_\mathrm w$. For a gigabyte interface and a packet size of
$1\,000\,$bytes, $q_\mathrm w = 62.5\,$packets.

All the experiments were done for different queue lengths, from the very short
size of just $B=25\,$ packets ($b=2.5$) to $B=350$ packets. It is important to
note that for the streamlined proposal, $B$ should be much greater than
$q_\mathrm w$, so experiments with queues smaller than $63\,$packets, although
represented for completeness, are not significant.

The first three figures represent the percentage of time spent active,
sleeping and transitioning for the three algorithms.

\begin{figure}
  \centering
  \includegraphics[width=\columnwidth]{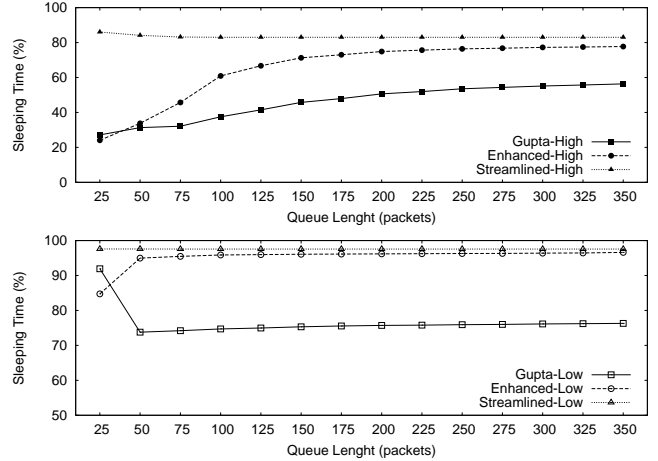}
  \caption{Total time in sleeping state.}
  \label{fig:off-time}
\end{figure}

\begin{figure}
  \centering
  \includegraphics[width=\columnwidth]{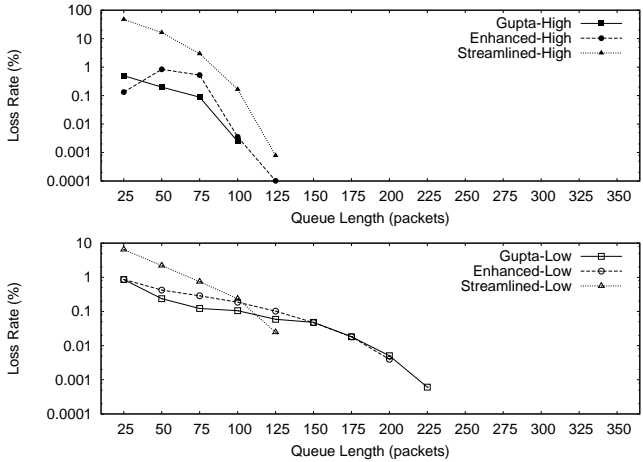}
  \caption{Packet dropped because of buffer overruns while sleeping.}
  \label{fig:drops}
\end{figure}

Fig.~\ref{fig:off-time} plots the total time spent in sleeping state.
The upper figure shows results for the high traffic trace, while the lower one
presents the results for the low traffic case. Results are similar, in any
case. The streamlined algorithm shows almost constant values for all buffer
sizes, however for very low buffer sizes this comes with the cost of very high
packet losses, as we will later show in Fig.~\ref{fig:drops}. Both Gupta-Singh
and our enhanced proposal get better results as the buffer, and thus $b$,
increases. This is expected as $t_\mathrm s$ depends in the size of $b$. The
difference between both methods come from the fact that the enhanced proposal
sleeps for longer periods as it waits for the queue to empty before sleeping.

The decision about when to start sleeping has dramatic consequences on the
amount of time spent transitioning to active.
\begin{figure}
  \centering
  \includegraphics[width=\columnwidth]{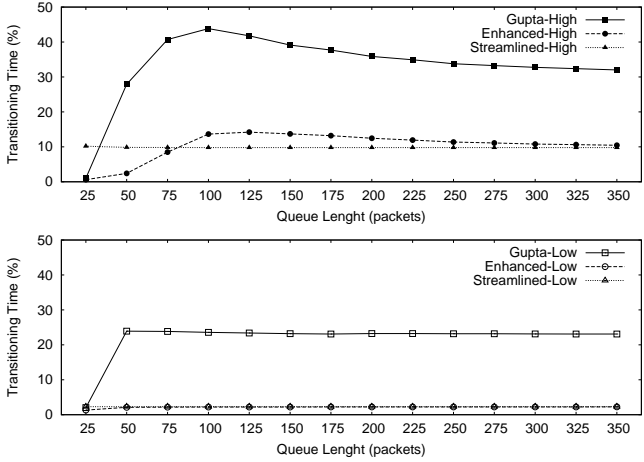}
  \caption{Total time in transitioning to active state.}
  \label{fig:tra-time}
\end{figure}
The longer the sleeping intervals, the less transitions. This can be clearly
seen in Fig.~\ref{fig:tra-time}. Just notice the big difference in the time in
transition for both the Gupta-Singh algorithm and the enhanced proposal. For
the first, the time in transition more than doubles our enhanced methods. This
has direct consequences in the energy savings, as the transitioning state is
very costly.

\begin{figure}
  \centering
  \includegraphics[width=\columnwidth]{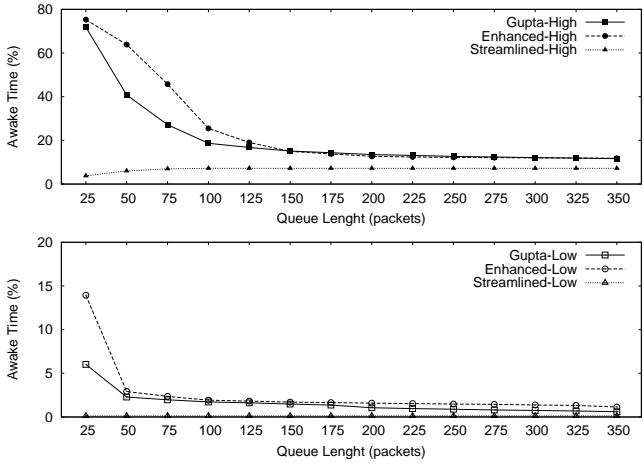}
  \caption{Total time in awake state.}
  \label{fig:on-time}
\end{figure}

Fig.~\ref{fig:on-time} shows the total time in active state.
This is similar for the three algorithms and very close to the real occupation
factor of the traffic traces. This is expected for our proposals, as the exit
from this state is based in a similar condition: the queue being empty. The
original Gupta-Sign is even more aggressive existing active state, but this
comes at the cost of shorter sleeping intervals and, correspondingly, much
more transitions.

\begin{figure}
  \centering
  \includegraphics[width=\columnwidth]{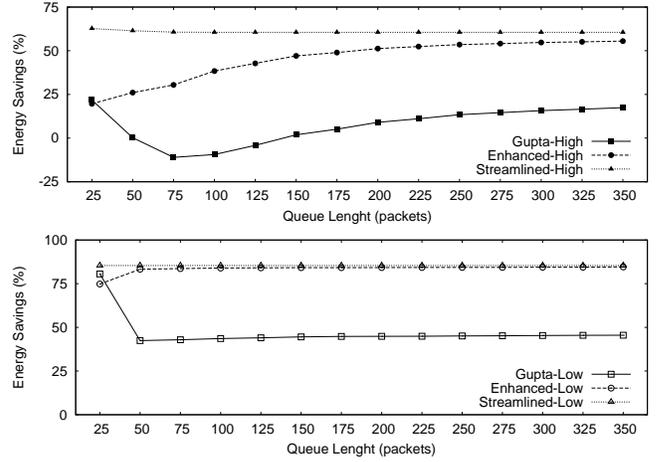}
  \caption{Energy savings for different buffer sizes.}
  \label{fig:savings}
\end{figure}

The real energy savings are depicted in Fig.~\ref{fig:savings}. This come from
comparison with the power a card that never enters sleeping state would
draw.\footnote{Although packet losses where not taken into account, their
  eventual retransmission can increase energy consumption. However, as we will
  show in later figures, packet drops caused by the sleeping algorithms are
  low enough to warrant any further consideration.}
Note how, in spite of its simplicity, the streamlined algorithm reaches the
highest savings of the three methods and is only comparable to the more
complex enhanced proposal. At the same time, it is important to note that the
Gupta-Singh algorithm can have pernicious effects on energy consumption in
some scenarios. For example, notice how for the high-traffic scenario for
small queue lengths it consumes more power than a non power-managed Ethernet
card.

\begin{figure}
  \centering
  \includegraphics[width=\columnwidth]{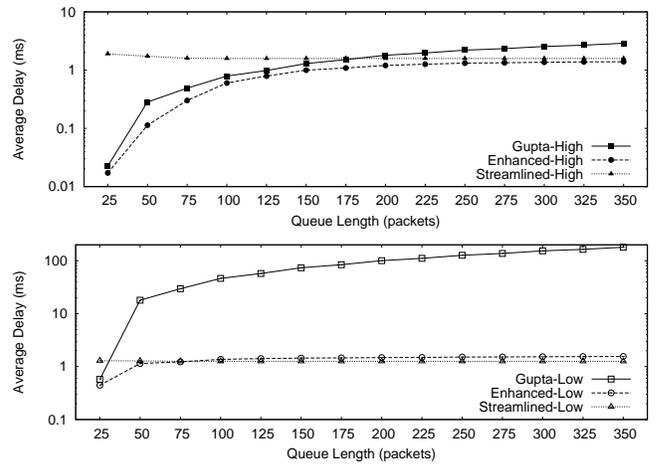}
  \caption{Average packet delay added by the sleeping algorithm.}
  \label{fig:delay}
\end{figure}

The last two figures show the performance cost that these power-management
methods have. Fig.~\ref{fig:delay} shows the effects on average packet delay.
As expected the delay increases with the buffer size. Note that for the
Gupta-Singh algorithm this delay can reach very high values if the conditions
for sleeping are favorable, that is, when there is very light traffic, as it
can enter the sleeping state even with traffic in the queue. For both enhanced
proposals this is not the case, as the sleeping interval is constrained to
$t_{\max}=2.5\,$ms and the sleeping interface is not reentered if there is
traffic in the queue. This limit is clearly identifiable in the figure.

Finally, Fig.~\ref{fig:drops} shows packet drops suffered because of the
sleeping methods. The streamlined proposal has very high-losses for very small
buffers, but this was expected as there is a minimum sensible value for $B \gg
q_\mathrm w = 62.5\,$packets. In fact, from $B > 1.5 q_\mathrm w \approx
100\,$packets, packet drops are comparable.
In any case, for sensible $B$ values packet drops are nearly negligible. In
fact we could not register losses with the sample traces for any buffer size
greater than $225\,$packets with neither method.

\section{Conclusions}
\label{sec:conclusions}

This paper provides new algorithm for exploiting low traffic load patterns
commonly found in Ethernet switches and endpoints. We first analyzed a well
known opportunistic sleeping algorithm
in~\cite{gupta07:_using_low_power_modes_for}. In it, its authors propose an
algorithm for shutting down transceivers so as to save power when there is low
load in an Ethernet link.

In our analysis of their proposal we found several shortcomings. Among those,
that the amount of introducing additional delay can be very high when the
traffic load is too small and, even more importantly, that energy savings are
not assured. In fact, under some circumstances the algorithm consumes more
energy than an Ethernet port running no power management algorithm at all.
This is due to too short sleeping intervals that draw more energy than the
energy saved while sleeping.

Based on our findings, we provided two alternative algorithms. One directly
based on the one in~\cite{gupta07:_using_low_power_modes_for} and a second one
that provides even slightly higher power savings with lower computational
complexity. We believe that a form of either of both proposals can be easily
implemented in Ethernet hardware for power savings of around $75\,\%$ with
respect of a non-power aware Ethernet card for typical workloads.

\section*{Acknowledgments}
\label{sec:acknowlgedgments}

We wish to thank Dr.~Singh from Portland State University for kindly providing
us with the traffic traces employed in his original study.

This work was supported by the ``Ministerio de Educación y Ciencia'' through
the project TSI2006-12507-C03-02 of the ``Plan Nacional de I+D+I'' (partly
financed with FEDER funds).

\balance{}

\bibliographystyle{IEEEtran}
\bibliography{IEEEabrv,biblio}

\begin{thebibliography}{10}
\providecommand{\url}[1]{#1}
\csname url@samestyle\endcsname
\providecommand{\newblock}{\relax}
\providecommand{\bibinfo}[2]{#2}
\providecommand{\BIBentrySTDinterwordspacing}{\spaceskip=0pt\relax}
\providecommand{\BIBentryALTinterwordstretchfactor}{4}
\providecommand{\BIBentryALTinterwordspacing}{\spaceskip=\fontdimen2\font plus
\BIBentryALTinterwordstretchfactor\fontdimen3\font minus
  \fontdimen4\font\relax}
\providecommand{\BIBforeignlanguage}[2]{{%
\expandafter\ifx\csname l@#1\endcsname\relax
\typeout{** WARNING: IEEEtran.bst: No hyphenation pattern has been}%
\typeout{** loaded for the language `#1'. Using the pattern for}%
\typeout{** the default language instead.}%
\else
\language=\csname l@#1\endcsname
\fi
#2}}
\providecommand{\BIBdecl}{\relax}
\BIBdecl

\bibitem{gupta03:_green_of_inter}
M.~Gupta and S.~Singh, ``Greening of the {I}nternet,'' in \emph{{SIGCOMM}'03:
  {P}roceedings of the 2003 conference on Applications, technologies,
  architectures, and protocols for computer communications}.\hskip 1em plus
  0.5em minus 0.4em\relax AMC, 2003, pp. 19--26.

\bibitem{ye02:_energ_effic_mac_protoc_for}
W.~Ye, J.~Heidemann, and D.~Estrin, ``An energy-efficient {MAC} protocol for
  wireless sensor networks,'' in \emph{Proceedings of the {IEEE} {INFOCOM}},
  vol.~3, 2002, pp. 1567--1576.

\bibitem{gupta07:_using_low_power_modes_for}
M.~Gupta and S.~Singh, ``Using low-power modes for energy conservation in
  {E}thernet {LAN}s,'' in \emph{Proceedings of the {IEEE} {INFOCOM}}, 2007, pp.
  2451--2455.

\bibitem{zhang08:_real_time_perfor_analy_of}
B.~Zhang, K.~Sabhanatarajan, A.~Gordon-Ross, and A.~George, ``Real-time
  performance analysis of adaptive link rate,'' in \emph{33rd {IEEE} Conference
  on {L}ocal {C}omputer {N}etworks, 2008. LCN 2008.}, Oct. 2008, pp. 282--288.

\bibitem{patel-predd08:_energ_effic_ether}
P.~Patel-Predd, ``Energy-efficient {E}thernet,'' \emph{{IEEE} Spectr.},
  vol.~45, p.~13, May 2008.

\bibitem{gupta04:_feasib_study_for_power_manag}
M.~Gupta, S.~Grover, and S.~Singh, ``A feasibility study for power management
  in {LAN} switches,'' in \emph{{IEEE} Internation Conference on Network
  Protocols}, 2004, pp. 1092--1648.

\bibitem{jardosh07:_towar_energ_star_wlan_infras}
A.~P. Jardosh, G.~Iannaccone, K.~Papagiannaki, and B.~Vinnakota, ``Towards an
  {E}nergy-{S}tar {WLAN} infrastructure,'' in \emph{Eighth {IEEE} Workshop on
  Mobile Computing Systems and Applications, 2007.}, Mar. 2007, pp. 85--90.

\bibitem{nedevschi08:_reduc_networ_energ_consum_via}
S.~Nedevschi, L.~Popa, G.~Iannaccone, S.~Ratnasamy, and D.~Wetheralll,
  ``Reducing network energy consumption via sleeping and rate-adaptation,'' in
  \emph{{NSDI}'08: Proceedings of the 5th {USENIX} Symposium on Networked
  Systems Design and Implementation}.\hskip 1em plus 0.5em minus 0.4em\relax
  Berkeley, CA, USA: {USENIX} Association, 2008, pp. 323--336.

\bibitem{06:_advan_config_and_power_inter_specif}
``Advanced configuration and power interface specification,''
  \url{http://acpi.info/spec.htm}, Oct. 2006.

\bibitem{karagiannis04:_nonst_poiss_view_of_inter_traff}
T.~Karagiannis, M.~Molle, M.~Faloutsos, and A.~Broido, ``A nonstationary
  {P}oisson view of {I}nternet traffic,'' in \emph{Proceedings of the {IEEE}
  {INFOCOM}}, vol.~3, 2004, pp. 1558--1569.

\bibitem{ns-2}
NS, ``{n}s {N}etwork {S}imulator,'' Mar. 2007,
  \url{http://www.isi.edu/nsman/ns/}.

\bibitem{08:_nortely}
``Nortel. converged data network solution,''
  \url{http://www.nortel.com/solutions/conv/collateral/tolly208298_nortel_conv%
erged_netenergy_costs_july08.pdf}, Jul. 2008.

\end{thebibliography}

\end{document}